\documentclass[10pt,sigconf,letterpaper,nonacm]{acmart}
\renewcommand\footnotetextcopyrightpermission[1]{} 
\usepackage{amsmath,amsfonts}
\usepackage{algorithmic}
\usepackage{graphicx}
\usepackage{textcomp}
\usepackage{xcolor}
\usepackage{siunitx}

\usepackage{subcaption}
\usepackage{multirow}
\usepackage{booktabs}
\usepackage{makecell}
\usepackage{tikz}

\newcommand{\solution}{\textsf{NetDiffus{}}}
\AtBeginDocument{%
  \providecommand\BibTeX{{%
    \normalfont B\kern-0.5em{\scshape i\kern-0.25em b}\kern-0.8em\TeX}}}

\setcopyright{acmcopyright}
\copyrightyear{2018}
\acmYear{2018}
\acmDOI{XXXXXXX.XXXXXXX}

\acmConference[Conference acronym 'XX]{Make sure to enter the correct
  conference title from your rights confirmation emai}{June 03--05,
  2018}{Woodstock, NY}
\acmPrice{15.00}
\acmISBN{978-1-4503-XXXX-X/18/06}

\begin{document}

\title{NetDiffus: Network Traffic Generation by Diffusion Models through Time-Series Imaging}


 \author{Nirhoshan Sivaroopan}

 \affiliation{
   \institution{University of Moratuwa}
   \country{Sri Lanka}
 }

 \author{Dumindu Bandara}
 \affiliation{
   \institution{University of Moratuwa}
   \country{Sri Lanka}}

 \author{Chamara Madarasingha}
 \affiliation{
   \institution{University of New South Wales}
   \country{Australia}
 }

 \author{Guilluame Jourjon}
 \affiliation{
  \institution{Data61-CSIRO}
  \country{Australia}}

 \author{Anura Jayasumana}
 \affiliation{
   \institution{Colorado State University}
   \country{USA}}

 \author{Kanchana Thilakarathna}
 \affiliation{
   \institution{The University of Sydney}
   \country{Australia}}

 \renewcommand{\shortauthors}{Nirhoshan, et al.}

\begin{abstract}

Network data analytics are now at the core of almost every networking solution. Nonetheless, limited access to networking data has been an enduring challenge due to many reasons including complexity of modern networks, commercial sensitivity, privacy and regulatory constraints. 
In this work, we explore how to leverage recent advancements in Diffusion Models (DM) to generate synthetic network traffic data. We develop an end-to-end framework - \solution{} that first converts one-dimensional time-series network traffic into two-dimensional images, and then synthesizes representative images for the original data. We demonstrate that \solution{} outperforms the state-of-the-art traffic generation methods based on Generative Adversarial Networks (GANs) by providing 66.4\% increase in fidelity of the generated data and 18.1\% increase in downstream machine learning tasks. We evaluate \solution{} on seven diverse traffic traces and show that utilizing synthetic data significantly improves traffic fingerprinting, anomaly detection and traffic classification. 

\end{abstract}




\maketitle

\section{Introduction}
Many network planning, monitoring, and optimization tasks depend on network data analytics in and off the network ~\cite{hussain2020machine,dimopoulos2016measuring}. 
These tasks are often driven by Machine Learning (ML) models which require a large amount of real network measurements to train the models. However, having access to appropriate network data traces is increasingly becoming challenging~\cite{lin2020using, yin2022practical, kattadige2021videotrain, shahid2020generative}. First, due to the complexity of modern networks and the sheer volume of data being transferred,  deploying data collection tools requires significant expertise and cost. Second, privacy and regulatory constraints have made many types of network data inaccessible or restricted in use for other purposes such as network management~\cite{kattadige2021videotrain, shahid2020generative}. Third, due to the commercial sensitivity of the data, many organizations do not share data with others, even among different departments of the same organization, let alone with the research community~\cite{lin2020using}.

To overcome these issues, \emph{synthetic data generation} has become a promising alternative. While there are many techniques and tools for data packets generation such as NS-3~\cite{ns3}, and iPerf~\cite{iperf} to satisfy a given model or a distribution, they fail to faithfully mimic the intricacies of real traces. However, recent ML-based approaches are capable of learning from traces to overcome this limitation~\cite{lin2020using, yin2022practical, kattadige2021videotrain, xu2019modeling,redvzovic1ip}. Among them, generative model based solutions such as DoppleGanger~\cite{lin2020using}, 
NetShare\cite{yin2022practical}, and 
CTGAN~\cite{xu2019modeling}, have shown superior performance in terms of representing practical network constraints and issues. 
However, GANs that have been the basis for many state-of-the-art (SOTA) network traffic generation tools suffer from mode-collapse, vanishing gradients, and instability unless the hyper-parameters are properly selected~\cite{dhariwal2021diffusion}. 

In this work, we explore how we can leverage recent advances in Diffusion Models (DM) architectures to generate synthetic network traffic. DM has shown outstanding performance compared to generative models such as GANs in particular with controlled image generation with models such as Dall-E~\cite{dalle2}. {The ability to control the generated output makes} DMs ideally suited for synthetic data generation for training ML models as it allows the generation of balanced datasets. This will lead to generating more robust and accurate trained models.
Nevertheless, DMs have not yet been investigated for network traffic generation.
\emph{To the best of our knowledge, this is the first attempt at utilizing Diffusion Models for network data generation}.

We propose \solution{}, a framework for network traffic generation using DM leveraging time-series imaging to achieve high fidelity  synthetic data.
First, we convert 1D network traces to a specific image format called Grammian Angular Summation Field (GASF)~\cite{wang2015imaging} to capture important features from 1D network traces. GASF images can encode features such as packet sizes, inter-packet times and most importantly the correlation among the 1D time-series samples onto an image in 2D space making it a rich source of information for ML models. Second, to reduce the computational demand and to improve the feature learning process in DMs, we apply several simple image processing techniques such as contrast adjustment and image resizing on GASF images. Finally, this enhanced data is used to train DMs and synthetic data from the trained models are used to improve various downstream ML tasks.


The goal of this work is to generate  network traffic features, either raw (e.g., packet size) or pre-processed (e.g, bytes downloaded by a group of packets~\cite{bronzino2019inferring}), in 2D GASF format and use it directly in improving downstream ML tasks~\cite{wang2015imaging, shankar2020epileptic, wan4176688novel, garibo2023gramian}. Note that, unlike recent works~\cite{lin2020using, yin2022practical}, \solution{} does not generate meta-data at this stage.

Leveraging a wide range of network traffic: video, web and IoT, we show that a standard DM can generate data with higher fidelity surpassing the baseline compared. For example, \solution{} achieves 28.0\% and 85.6\% fidelity improvement compared to SOTA GAN based models, DoppelGanger~\cite{lin2020using} and NetShare~\cite{yin2022practical} respectively. Moreover, we utilize these synthetic data to train ML models related to different network-related tasks, such as traffic fingerprinting, anomaly detection, and classifications in data-limited scenarios. Even without combining with original data, \solution{} can achieve almost the same accuracy of original data or improved accuracy of 1--57\% in those tasks. Comparing with the baselines above, \solution{} synthetic data can improve classification performance by 4.7--32.3\% in the corresponding ML tasks. Artifacts  are --\textit{hidden for double blind review--.}


\section{Background and Related work}

\textbf{Synthetic network traffic generation: }
A Plethora of work has been done in data generation domain~\cite{yin2022practical, lin2020using, kattadige2021videotrain,goodfellow2020generative,croitoru2023diffusion}.
Markov models and recurrent neural networks are commonly utilized in prior network traffic generation models~\cite{redvzovic1ip,lin2020using}. Despite the generalization they provide, their fidelity in domain-specific generative tasks remains limited. 
In the recent past, Generative Adversarial Network (GAN) based models have revolutionized network traffic synthesis, offering powerful capabilities. Generator and discriminator architecture used in GANs  can effectively extract the characteristics of network traces with further modification to preserve temporal attributes~\cite{goodfellow2020generative,lin2020using, yin2022practical, kattadige2021videotrain, xu2019modeling}. Despite their promise, GANs suffer from major issues such as mode collapsing, and unstable and inflexible training~\cite{lin2020using,dhariwal2021diffusion}.


\noindent
\textbf{{Imaging Time Series Data: }}
Converting 1D data into 2D images is widely studied in many works~\cite{wang2015imaging,said2019deep,fukino2016coarse,smith2020conditional,smith2021spectral}. One motivation for such conversion is the improved performance in downstream analysis tasks, especially in ML-based classification~\cite{wang2015imaging}. Also, such image representations  are rich with information for ML tasks~\cite{smith2021spectral}. 
Using Gramian matrix and Markov Transition Fields (MTF), the authors in~\cite{wang2015imaging} have converted 1D data into specific images formats called Gramian Angular Summation Field (GASF)  and Gramian Angular Difference Field (GADF). 
These data formats are derivatives of Gramian Angular Field (GAF), generated by converting the time series into a polar coordinate system and mapping the correlation between the 1D samples.

\noindent
\textbf{{DM and its promise in data generation}: }
DM falls into likelihood-based methods which have more distribution coverage, scalability and stability in training and is a solution for issues such as mode collapse, instability and less flexibility in GAN models~\cite{dhariwal2021diffusion, guarnera2023level, rombach2022high}. In the forward pass, DM gradually adds Gaussian noise to the input image until it becomes pure noise. Then a DNN model is trained to denoise the image to recover the original image. This trained DNN acts as a generative model which produces images from pure noise distributions. Many recent works have used DMs in tasks such as  image, audio, text-to-image and image-to-text generation, nevertheless have not been used in network traffic generation with various network traffic types~\cite{dhariwal2021diffusion,corvi2023detection,zhang2022motiondiffuse,croitoru2023diffusion}.

\textit{In contrast to the previous work, we demonstrate how DMs can be used for network traffic generation by converting time series distribution to 2D GASF images. These images accurately capture feature distributions including correlations between 1D sample points further supporting downstream ML tasks.}

\section{Design of \solution{}}\label{sec:design}



\begin{figure*}[h]
    \centering
    \begin{subfigure}{.26\textwidth}
        \centering
        \includegraphics[width=\linewidth]{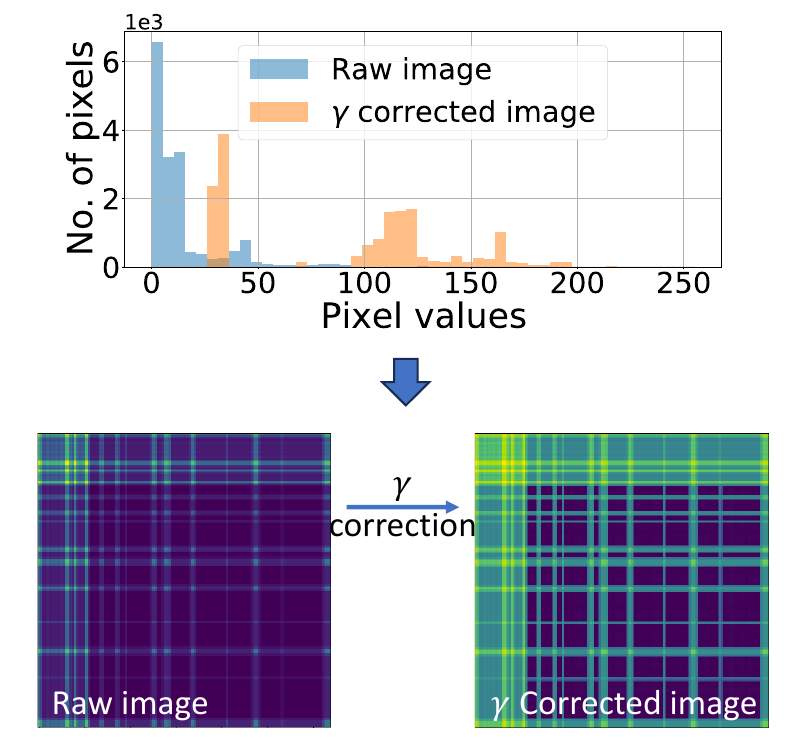}
        \vspace{-6mm}
        \caption{Histogram distribution between raw and $\gamma$ corrected image}
        \label{fig:gamma_correction}
    \end{subfigure}
    \hfill
    \begin{subfigure}{.65\textwidth}
        \centering
        \includegraphics[width=\linewidth]{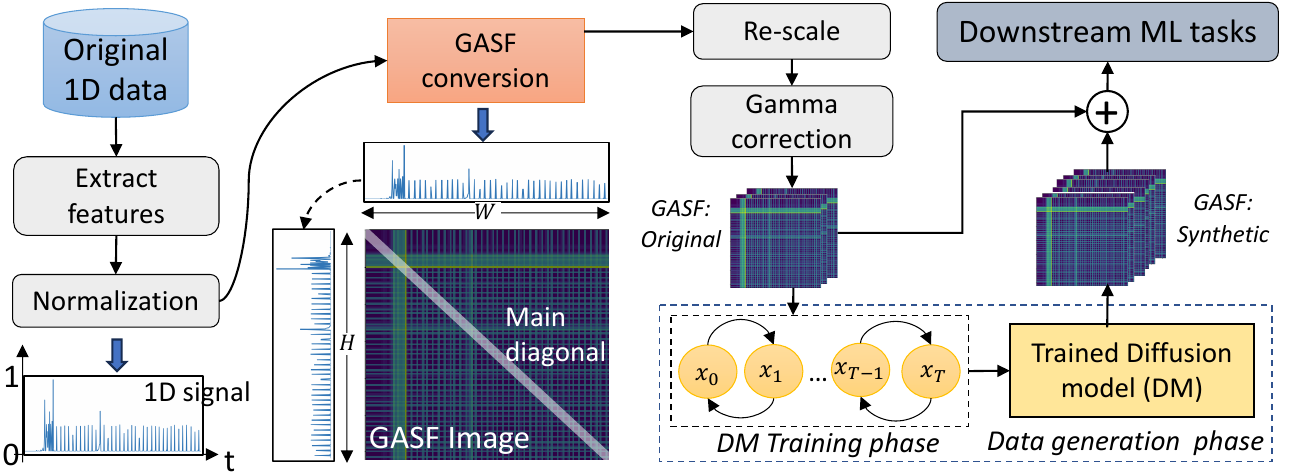}
        \vspace{-6mm}
        \caption{Overall design  of \solution{} including sample 1D signal and a GASF image and the key design steps in described in Section~\ref{sec:design}}
        \label{fig:overview}
    \end{subfigure}
\vspace{-4mm}
\caption{Impact of gamma correction taking sample GASF image and overall design of \solution{}}\vspace{-4mm}
\label{fig:fidelity}
\end{figure*}

\subsection{Capturing important feature attributes}\label{subsec:capture_important_feature}
DNNs are designed to learn hidden features of input data. However, identifying subtle features such as correlation between the samples, frequency-related patterns from 1D signals or time series, etc., requires complex models and  rigorous training processes. Manually extracting such features prior to model training is  non-trivial as it can enable the model to learn those features efficiently and improve the data fidelity. To achieve this, in \solution{}, we convert 1D signals into 2D image format, GASF, following the method in~\cite{wang2015imaging}. GASF images map features such as amplitudes, inter-packet gaps and temporal correlations on to one 2D space.

Given a 1D signal of a network feature (e.g., bytes dl (downloaded)), we first convert it to polar coordinates followed by creating the corresponding Grammian matrix. Here, elements of the Grammian matrix denote the inner product between the cosine angles of the samples from the time series in the polar coordinate system, which is the base of the GAF. By taking the inner product, the matrix further represents a correlation map between the sample points in the 1D trace~\cite{wang2015imaging}. Then, we create  GASF images by taking the summation of  all the  pairs of elements in row and column directions to  remove the dependency on the radius in  GAF data. Appendix~\ref{append:gasf conversion} further explains GASF conversion.  Fig.~\ref{fig:overview} shows a sample GASF image.
The width ($W$) and height ($H$) of the image are equal to the trace length. The main diagonal of the image corresponds to the  time-series signal and contains encoded feature amplitudes, inter-packet gaps, etc. 

\subsection{Highlighting hidden features}

As we operate in 2D domain, enhancing the contrasts of GASF images can further highlight subtle feature variations that can be effectively learnt by DMs and improve the fidelity. We leverage standard gamma correction on raw GASF images according to the equation, $I_c = A*I^{\gamma}_r$, where $I_r$, $I_c$, $A$ and $\gamma$ are gamma corrected image, raw image, a constant and gamma variable respectively. We empirically set $\gamma=0.25$ and $A=1$. Fig.~\ref{fig:gamma_correction} shows the histogram distribution of sample raw and gamma-corrected images. We notice that this process separates the pixel values into distinct ranges increasing the image contrasts and emphasizing the feature variations.


\subsection{Supporting fast and stable training}

DMs typically require high computational power and time. Hence, keeping the GASF image size similar to the trace length can lead to a longer training time and insufficient resources. As a solution,  we resize the images to a fixed smaller resolution and feed low resolution images for DM training. We use \texttt{OpenCV.resize()} method with \texttt{INTER\textunderscore AREA} interpolation method which resamples image pixels based on area relations and is the preferred method for image decimation~\cite{opencv_resize}. 
We empirically decide the image size without affecting the downstream ML performance as image resizing can potentially drop high-frequency information from the images. Also, we max normalize GASF pixel range to [0,1] dividing each pixel by the  global maximum value, 255. This makes DM training process faster and more stable. Note that we leverage vectorized operations in \texttt{Python-numpy}  to speed up these pixel level operations.


\subsection{Overall design of \solution{}}

Fig.~\ref{fig:overview} presents the overall process of \solution{}. We start with time-series feature extraction from related datasets and max normalization. Then, the 1D signals are converted to GASF images which are further enhanced by gamma conversion and resizing the images. Finally, we use these original GASF images to train DMs. Unless otherwise noted, from each dataset, we use the first 80\% of the data for synthetic data generation and keep the remaining as the test dataset for the downstream ML tasks. We set diffusion steps to 1000 and a standard U-Net model with 5 layers for the denoising process. The synthesized GASF images are used to improve various downstream ML tasks combined with original GASF data. We will release all the model details with the artefacts.

Interestingly,  we observe that a basic DM architecture suffices to  generate GASF images with high fidelity. 
At this stage of \solution{}, we do not construct the corresponding 1D traces from the GASF images for downstream analysis, on the one hand, for a variety of ML-based analysis, a 2D image is a suitable format~\cite{wang2015imaging, shankar2020epileptic, wan4176688novel, garibo2023gramian}.  On the other hand, we observed improved ML classification performance with 2D GASF data  compared to its 1D counterpart as described in Appendix~\ref{append:1d vs 2d original}.  However, reconstruction of 1D traces can be easily done by applying the Equation: $Trace = \frac{\sqrt{Y+1}}{2}$. Here $Y$ is the vector consisting of the elements in the main diagonal of GASF image~\cite{wang2015imaging,shankar2020epileptic}. We keep further evaluations with reconstructed 1D data in our future work.

\vspace{-2mm}\section{Evaluation and results}\label{sec:eval and result}


\subsection{Setup}
\subsubsection{Dataset}\label{subsubsec:dataset}
We collected two main datasets (\textbf{D1)} and \textbf{D3}) and selected one publicly available dataset (\textbf{D2}) to address diverse conditions present in networks.

\noindent
\textit{\textbf{D1: Streaming videos:}}~We selected videos from \textbf{YouTube (YT)}, \textbf{Stan} and \textbf{Netflix}, 20 from each with 3 min duration, and streamed each video multiple times to generate 100 traces. While streaming, we passively captured network packets which are binned into non-overlapping 0.25~s bins to extract \textit{Total bytes dl} feature in each bin. Binning can highlight different video-specific features (e.g., quality switching) in network traces and increase the performance in downstream ML tasks~\cite{kattadige2021360norvic,bronzino2019inferring}. 
The intended ML task is to fingerprint a given trace into one of the 20 videos from a given platform. 

\noindent
\textit{\textbf{D2: Accessing web pages:}}~ We selected publicly available web surfing dataset from~\cite{sirinam2018deep} which has  20 websites. The extracted features include packet direction (i.e., (+1) for uplink and (-1) for downlink packets), inter-packet gaps. Each trace has 5000 fixed number of samples. Traces with over 5000 samples, it is truncated and otherwise 0 padded up to 5000 samples. Each traffic trace is classified into one of the 20 classes (i.e., website) as a website fingerprinting task. 

\noindent
\textit{\textbf{D3: Traffic generated by IoT smart-home devices}}~We collected this dataset by passively monitoring the network traffic from two smart home assistance devices \textbf{{Google Home}}~\cite{online_google} and \textbf{Amazon alexa}~\cite{online_alexa} smart home devices. A user gives 10 different commands for each device and the device communicates with its cloud server to execute the related activity. By repeating each command, we collected 1000 traces and each trace is kept at 300 packet length following the same trace truncation and 0 padding approach in \textbf{D2}. Captured features include packet size, direction and inter-packet gaps. The  ML task is to classify each trace into one of the activities. 

The datasets we selected consist of a wide range of features, for example from raw packet sizes to aggregated \textit{Total bytes dl }values by bins that can be used for a wide range of ML tasks. This further verifies the robustness of \solution{} for different network-related feature generation along with the efficacy of GASF-based synthetic data for downstream tasks.
Based on how we utilize the data for ML training, there are three main scenarios; \textit{i})~\textbf{original}: use only original data, \textit{ii})~\textbf{synth}: use only synthetic data, \textit{iii})~\textbf{ori+synth}: combine original data with synthetic. Unless otherwise noted, we separate data (i.e., network traffic traces) from each class from each dataset  into 80\%-20\% train - test splits which are used to train and test both DMs in data generation and  ML models in downstream tasks.

\vspace{-2mm}
\subsubsection{Benchmark models}

\textbf{DoppelGanger (DG)~\cite{lin2020using}:}~A GAN based approach which generates both metadata and traffic features of the traces while finding their correlations. 
We train \solution{} using one of their datasets, Wikipedia Web Traffic (WW), first, to compare  DG with \solution{} while preserving its original attributes, and second,  to show the robustness of \solution{} to different datasets.\\
\textbf{NetShare~\cite{yin2022practical}:}~SOTA GAN-based method for packet/flow header generation taking them as time-series data compared to tabular format.
Though the base model considered is DG, the authors claim that with the proposed packet/flow data epochs merging mechanism, the scalability of the generation has been increased with improved fidelity in synthetic data. We train Netshare using \textbf{D3-Google} and \textbf{D3-Alexa} datasets as these datasets are compatible with packet-level data generation in Netshare.

\textbf{\textit{Note:}} DG and NetShare have outperformed many other ML based and statistical approaches~\cite{esteban2017real,yoon2019time,cheng2019pac,xu2020stan}, and therefore, we exclude other GAN methods and ML based approaches from the comparison. Since these models synthesize 1D data, we convert 1D synthetic traces from those models to GASF images before the comparison.


    

\subsection{Analysis of data fidelity}

\begin{figure}[t]
    \centering
    \begin{subfigure}{.49\columnwidth}
        \centering
        \includegraphics[width=\linewidth]{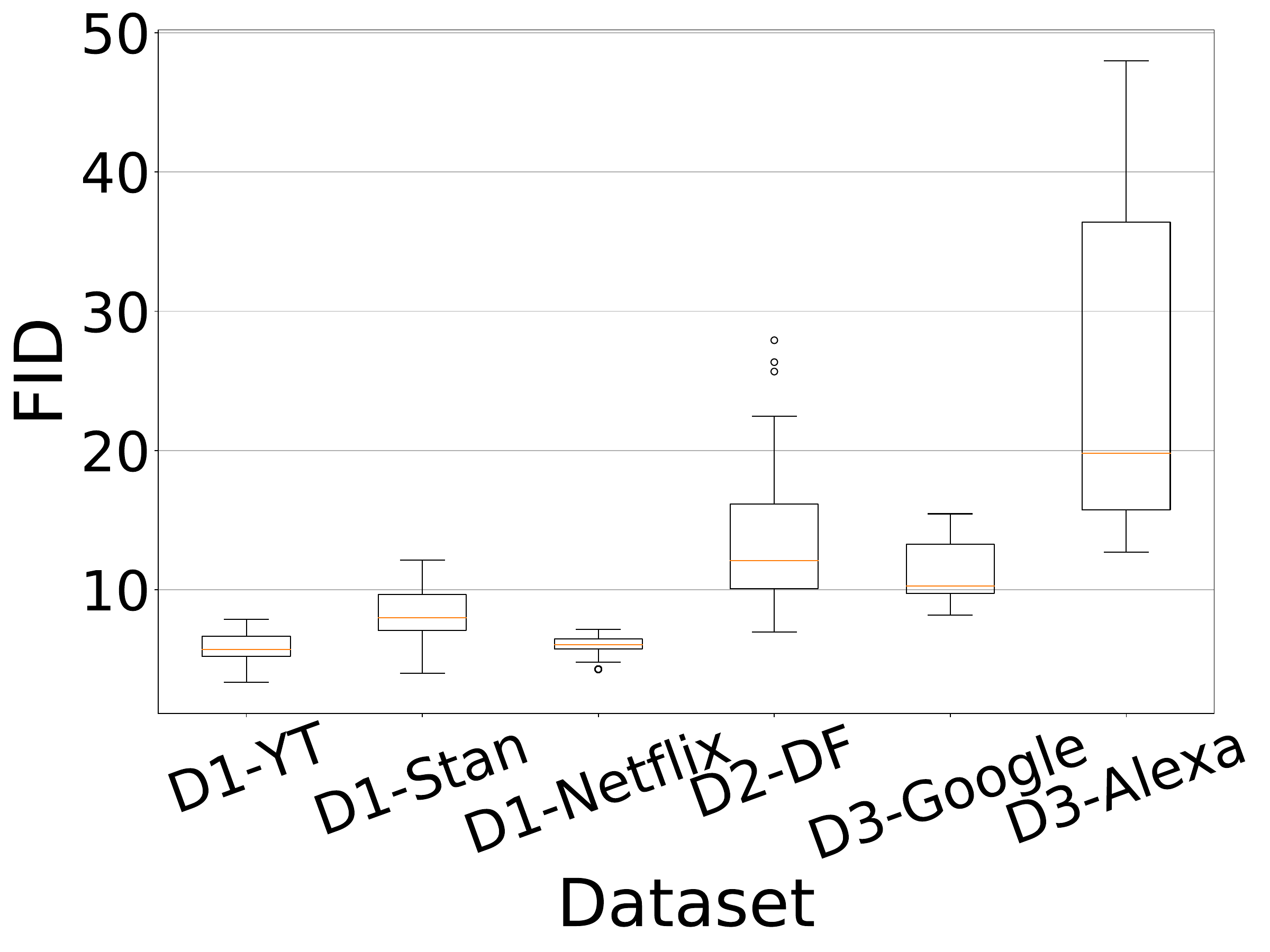}
        \vspace{-6mm}
        \caption{FID score by datasets}
        \label{fig:fid}
    \end{subfigure}
    \hfill
    \begin{subfigure}{.18\columnwidth}
        \centering
        \includegraphics[width=\linewidth]{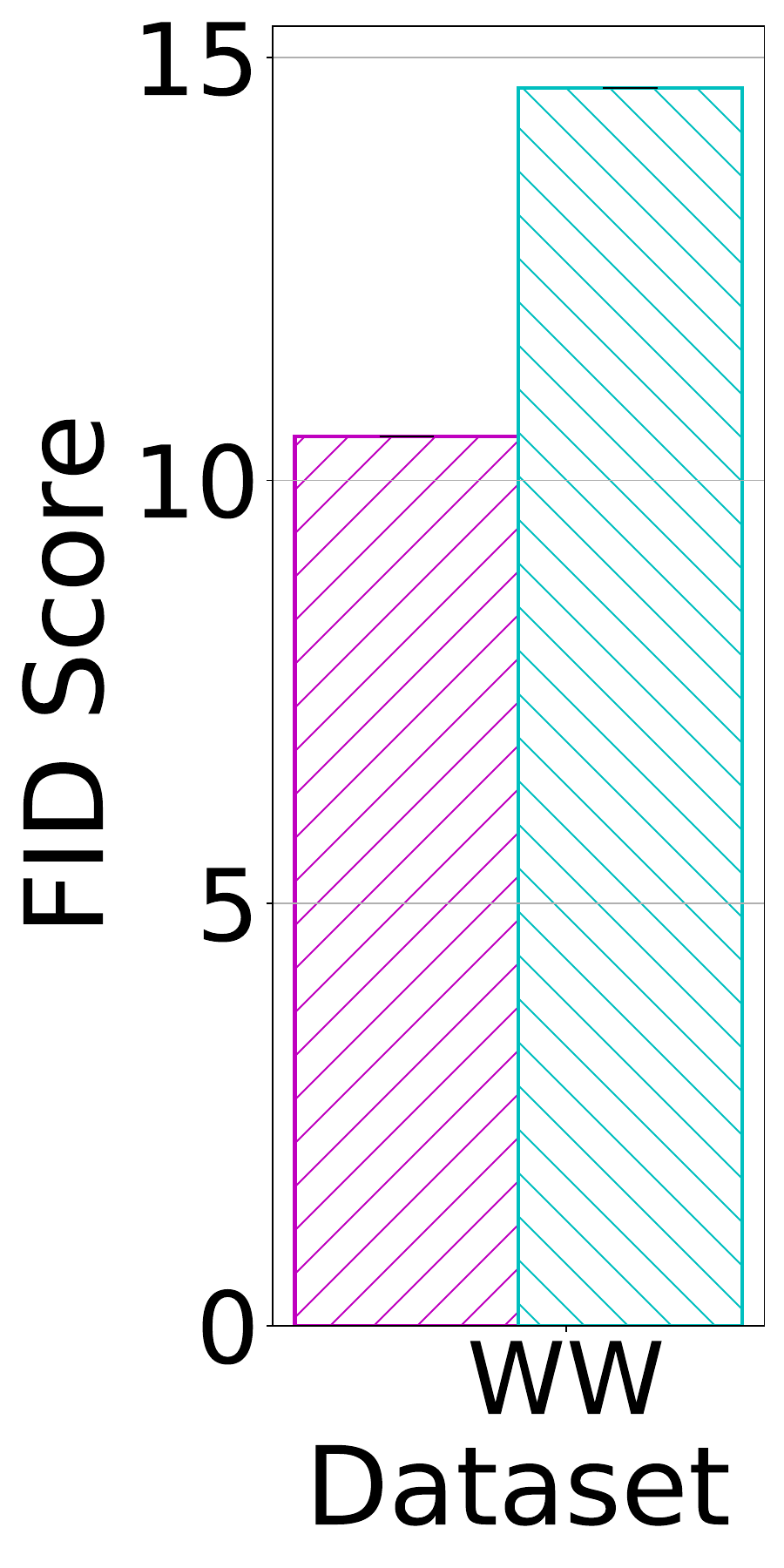}
        \vspace{-6mm}
        \caption{DG}
        \label{fig:fid-baseline-DG}
    \end{subfigure}
    \hfill
    \begin{subfigure}{.30\columnwidth}
        \centering
        \includegraphics[width=\linewidth]{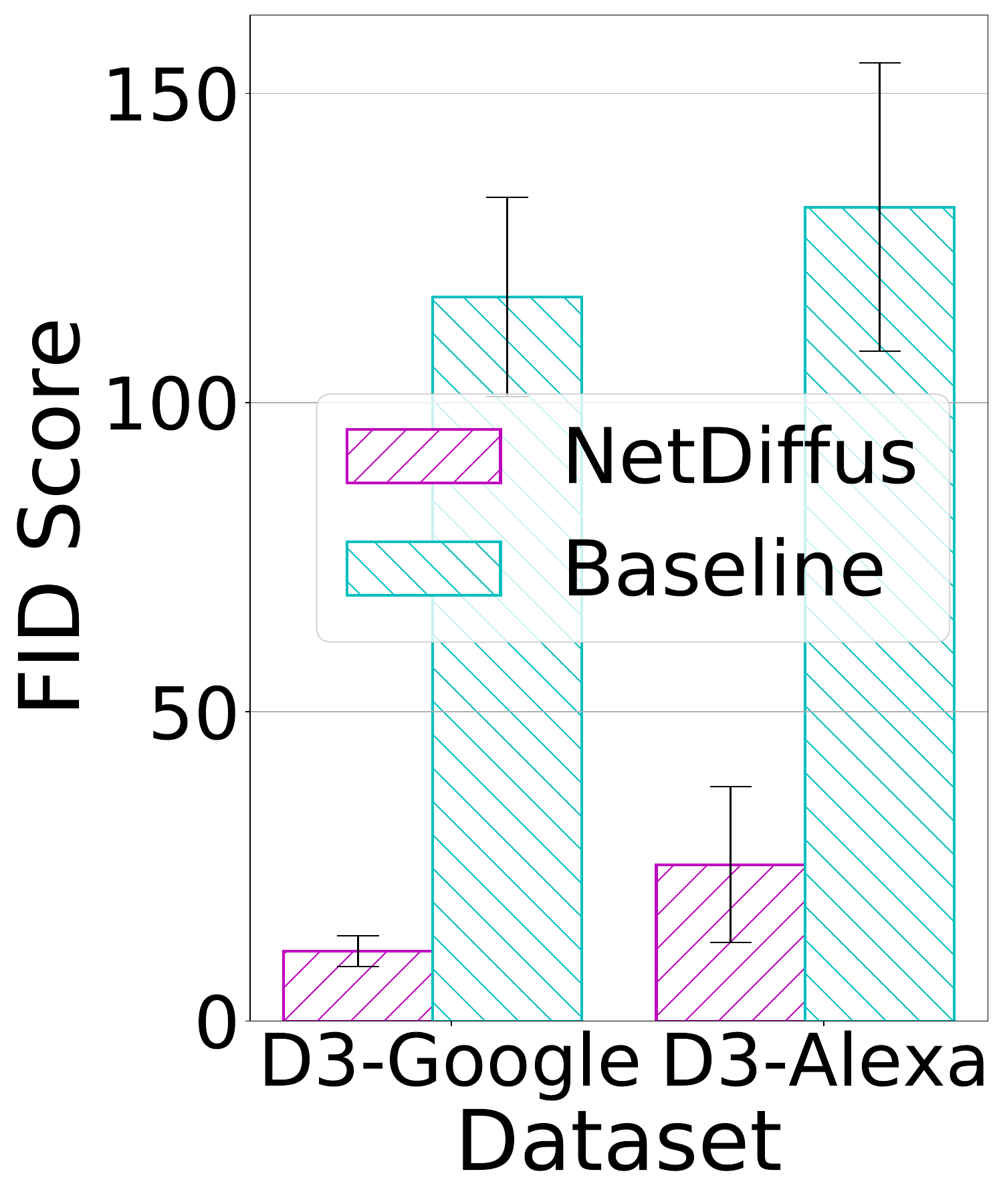}
        \vspace{-6mm}
        \caption{Netshare}
        \label{fig:fid-baseline-Netshare}
    \end{subfigure}    
\vspace{-4mm}
\caption{Fidelity by individual datasets and comparison with baselines.}\vspace{-4mm}
\label{fig:fidelity}
\end{figure}

We use the metric FID (Frechet Inception Distance)~\cite{chong2020effectively} to evaluate fidelity in synthetic traces.\footnote{FID is a metric used to measure fidelity in images. FID values $<20$ are commonly observed for high fidelity data~\cite{song2020improved,ravuri2019classification,chong2020effectively}.}
A lower FID score means that the original and synthetic images have a close distribution and vice-versa. 
In these experiments, we  pair-wise compare $n$ randomly selected   synthetic traces from each class with corresponding original traces used to train the model. Fig.~\ref{fig:fid} shows that, overall, \textbf{D1} data has a lower FID score compared to others which is less than 9 on average. \textbf{D3--Alexa}, shows the highest FID score, 25.8 ($\pm$11.9), because of the overlap in the features between classes for some traces in the original dataset.
We further analyze the histogram distribution between original and synthetic data in Appendix~\ref{append:histo d3}.

Fig.~\ref{fig:fid-baseline-DG} and Fig.~\ref{fig:fid-baseline-Netshare} compares the FID scores between \solution{} and the baselines: DG and Netshare respectively. Overall, \solution{} outperforms all the baselines by an average FID value difference of 72.8 which is equivalent to 66.4\% gain.
\solution{} surpasses DG by 4.1 (i.e., a gain of 28\%) showing its applicability to new datasets, and the model significantly outperforms Netshare by around 105 of average FID score difference (i.e., a gain of 85.6\%). The main reason for lower fidelity in Netshare is that the model is not able to learn packet size distributions  and correlations over a wide range (e.g., 0-1500 bytes) despite the model robustness claimed~\cite{yin2022practical}.  Though we do not evaluate \solution{} fidelity in 1D domain after reversing GASF conversion, the higher fidelity in 2D GASF format indicates the high fidelity in 1D domain as well.

\subsection{Performance in downstream ML models}

\subsubsection{ML accuracy comparison with baselines:}Fig.~\ref{fig:ml-dg} compares the \solution{} with DG model using five classification ML algorithms (Convolutional Neural Network (CNN), XGBoost, Multi-Layer Perceptron (MLP), Naive Bayes (NB), Random Forest (RF)). We classify the type of access to Wiki pages, e.g. mobile, desktop, etc. in WW trace for \textbf{original} and \textbf{synth} scenarios taking an equal number of synthetic samples from \solution{} and DG. In all ML models, \solution{} exceeds the DG accuracy, which is 4.67\% on average. Similarly, in Fig.~\ref{fig:ml-netshare}, for both \textbf{D3-Google} and \textbf{D3-Alexa} datasets, \solution{} outperforms Netshare by an average difference of 32.3\% and 17.3\% respectively. These results indicates that \solution{} can outperform many SOTA data generation models.

\subsubsection{Performance in different downstream ML models:}
\begin{figure}[t]
\centering
\begin{subfigure}{.48\columnwidth}
  \centering
  \includegraphics[width=\linewidth]{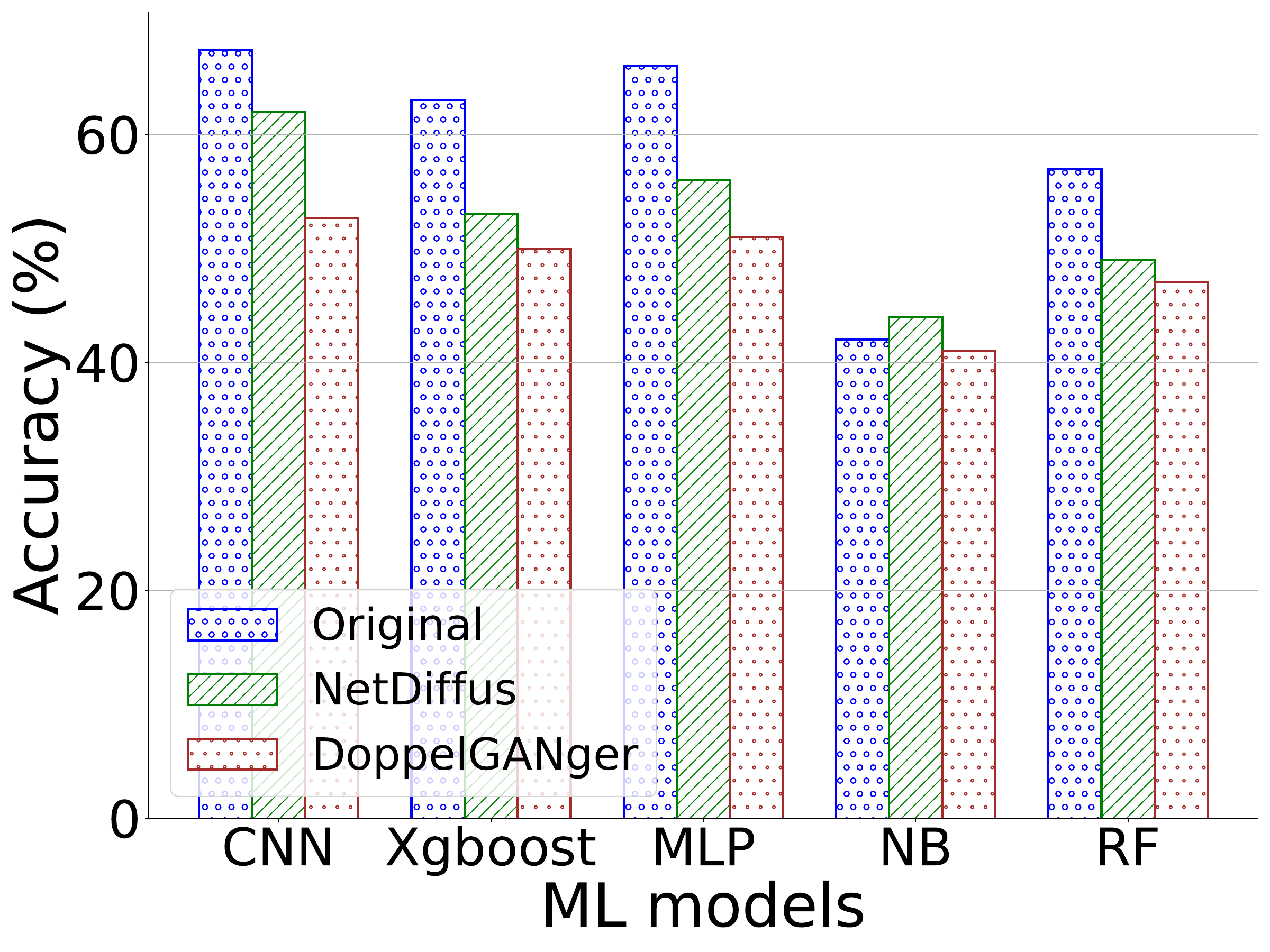}
  \vspace{-6mm}
  \caption{DG: WW dataset}
  \label{fig:ml-dg}
\end{subfigure}%
\hfill
\begin{subfigure}{.48\columnwidth}
  \centering
  \includegraphics[width=\linewidth]{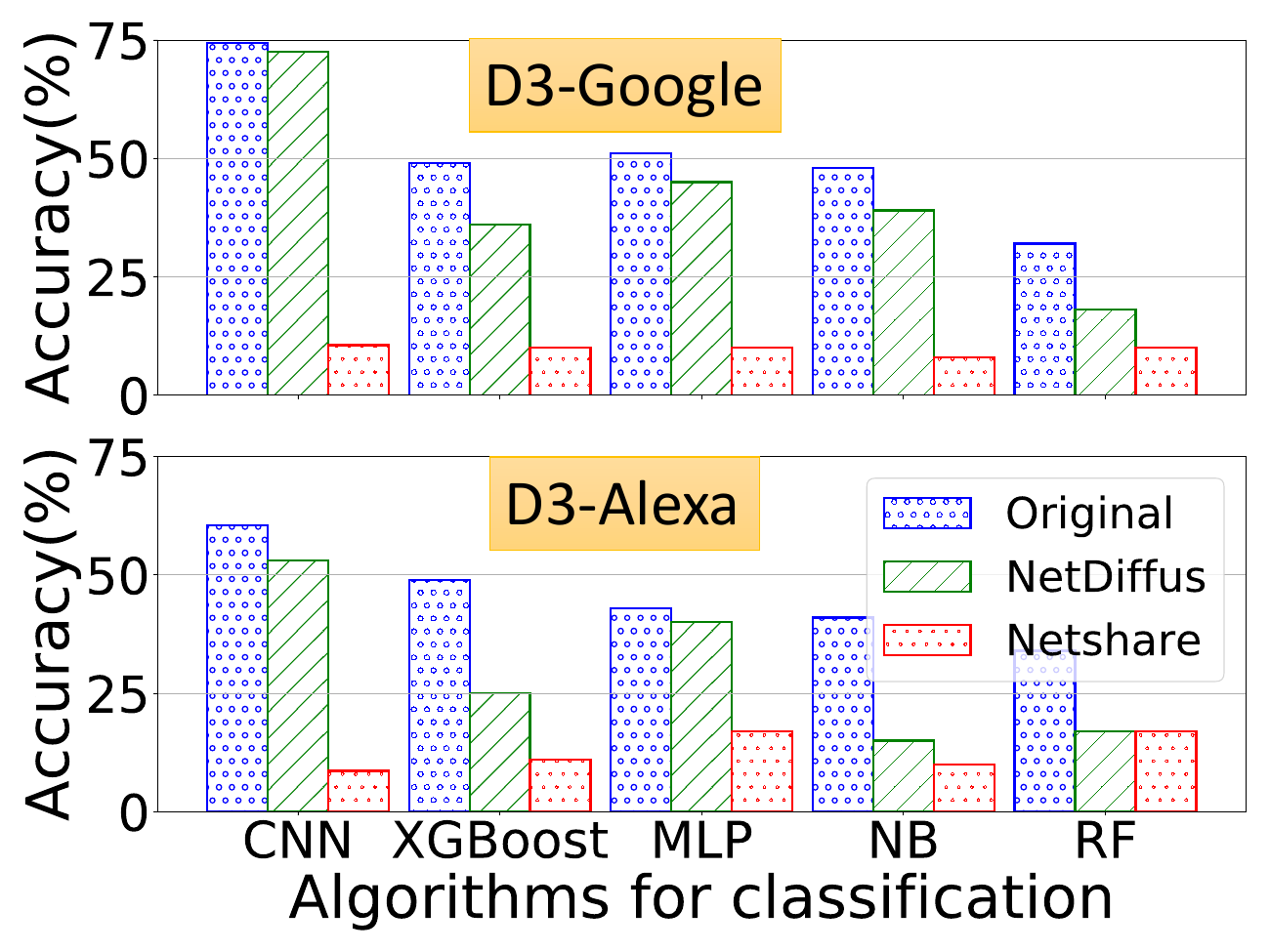}
  \vspace{-6mm}
  \caption{Netshare: D3-G \& D3-A }
  \label{fig:ml-netshare}
\end{subfigure}
\vspace{-4mm}
\caption{Comparison with baselines G:Google, A:Alexa\vspace{-4mm}}
\label{fig:ml-acc-compare}
\end{figure}

By considering multiple downstream ML algorithms in Fig.~\ref{fig:ml-acc-compare}, we further show that \solution{} synthetic data can be used to evaluate different downstream ML algorithms as well. This is important when utilizing synthetic data to tune models in load balancing, cluster scheduling etc.~\cite{lin2020using}. A key property of synthetic data to achieve this goal is they should have accuracy trends similar to that of original data in different algorithms. Except for the  \textbf{D3-Alexa}-MLP evaluation in Fig.~\ref{fig:ml-netshare}-bottom, \solution{} synthetic data follows similar accuracy patterns to its original data in all other cases. For example, in Fig.~\ref{fig:ml-dg}, CNN, XGBoost and MLP show higher classification accuracy with both original and \solution{} data and both datasets show lower accuracy with NB.

\subsection{Improved ML performance in use-cases}


\subsubsection{Surveillance through traffic fingerprinting}\label{subsubsec:hie-clf}

ML has been widely applied for the surveillance of network traffic, nonetheless, it can show limited performance due to the shortage of  training data. In this use-case, we analyze how traffic fingerprinting tasks can be optimized by \solution{} synthetic data. We leverage a hierarchical ML classifier under three types of classification :~L1: \textit{Traffic type} (e.g., video, web or IoT), L2: \textit{Platform type} (e.g., YT, Stan), and L3: \textit{Individual classes} (e.g., individual video, website).  

\noindent
\textit{\textbf{{Results and observation:}}}
Table~\ref{table:hierarchical clf} reports classification accuracy. We observe both L1 and L2 provide over 95\% accuracy in both \textbf{original} and \textbf{synth} scenarios. L3 is a challenging task compared to L1 and L2 due to the higher number of classes and the similarities in traces we notice between the classes. In \textbf{D1}, video fingerprinting task we see 3.5--6.5\% accuracy improvement in \textbf{synth} scenario compared to \textbf{original} data mainly due to the high fidelity in synthetic data 
{Though we see 5.83\%  average accuracy drop in \textbf{D2} and \textbf{D3} in \textbf{synth} scenarios compared to \textbf{original}, referring to recent literature~\cite{lin2020using, yin2022practical} and considering the  difficulty in tasks, we believe such accuracy levels are still acceptable}. However, by combining original with synthetic data, we achieve improved accuracy compared to \textbf{original} scenario by 1--8\%.

\begin{table}[t]
    \small
    \centering
    \caption{Accuracy of hierarchical classification model}
    \vspace{-4mm}
    \label{table:hierarchical clf}
    
    \begin{tabular}{llccc}
        \hline
        \textbf{Layer (type)}  & \textbf{Data used}      & \textbf{original} & \textbf{synth} &\textbf{ori+synth}\\
        \hline
        L1 (\textit{Traffic})&\textbf{D1}+\textbf{D2}+\textbf{D3} & 99.0    & 100.0 & 100.0 \\
        \hline
        L2 (\textit{Platform})&\textbf{D1}    & 100.0    & 100.0 & 100.0 \\
         &\textbf{D3} & 97.0    & 96.5 & 98.5 \\
         \hline
         L3 (\textit{Class}) 
        
         &\textbf{D1-YT} & 84.5    & 91.0 & 92.5 \\
         &\textbf{D1-Stan} & 92.5    & 98.5  & 99.5\\
         &\textbf{D1-Netflix}    & 97.5    & 100.0 & 100.0\\
         &\textbf{D2-DF} & 92.0    & 83.8 & 93.64  \\
         &\textbf{D3-Google} & 74.4    & 72.5 & 77.0 \\
         &\textbf{D3-Alexa} & 60.5    & 53.1 & 62.0 \\
         
        \hline
    \end{tabular}
\end{table}

\begin{figure}[t]
\centering
\begin{subfigure}{.48\columnwidth}
  \centering
  \includegraphics[width=\linewidth]{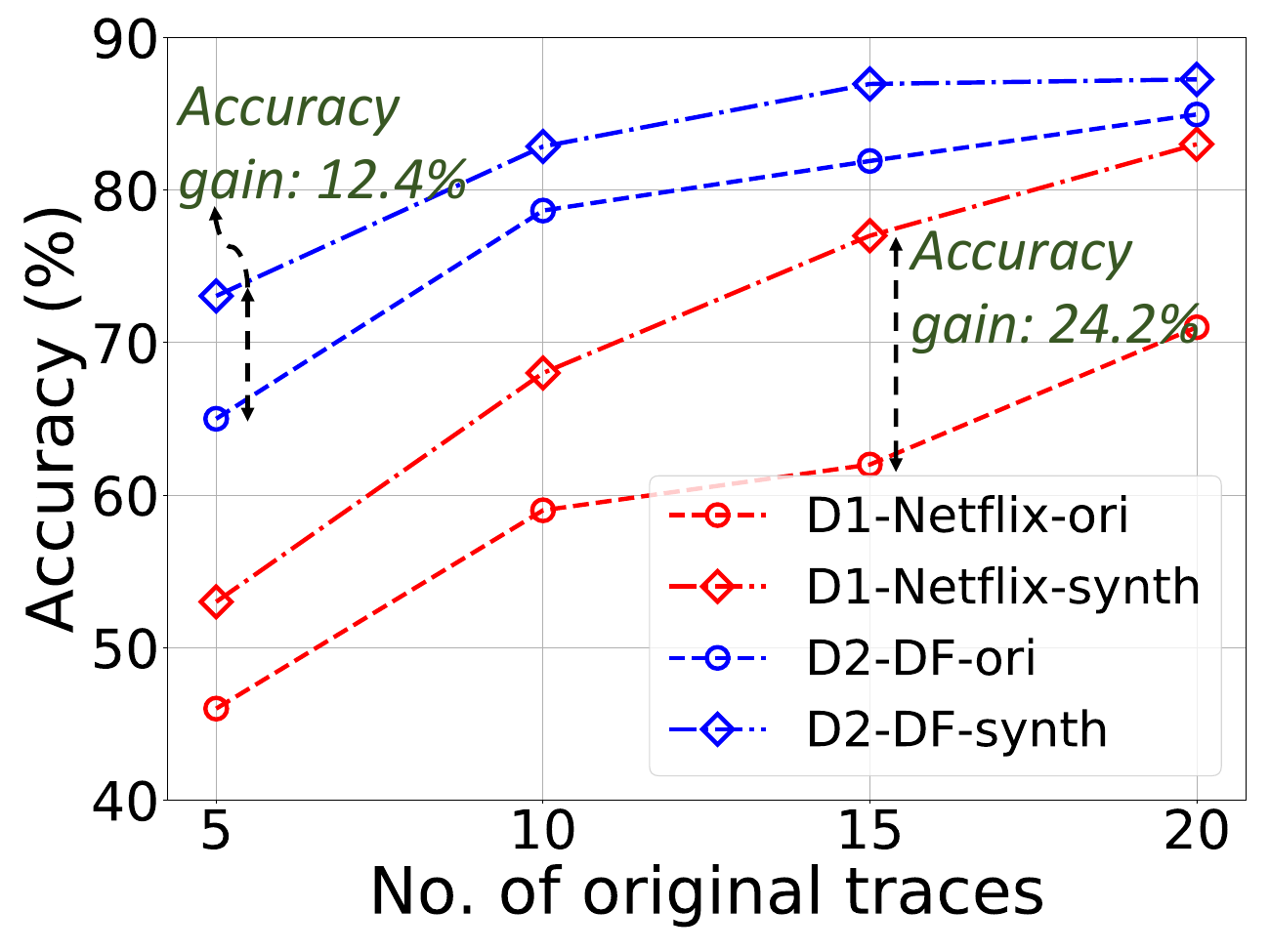}
  \vspace{-6mm}
  \caption{\textbf{D1-Netflix} and \textbf{D2-DF}}
  \label{fig:netflix-df}
\end{subfigure}%
\hfill
\begin{subfigure}{.48\columnwidth}
  \centering
  \includegraphics[width=\linewidth]{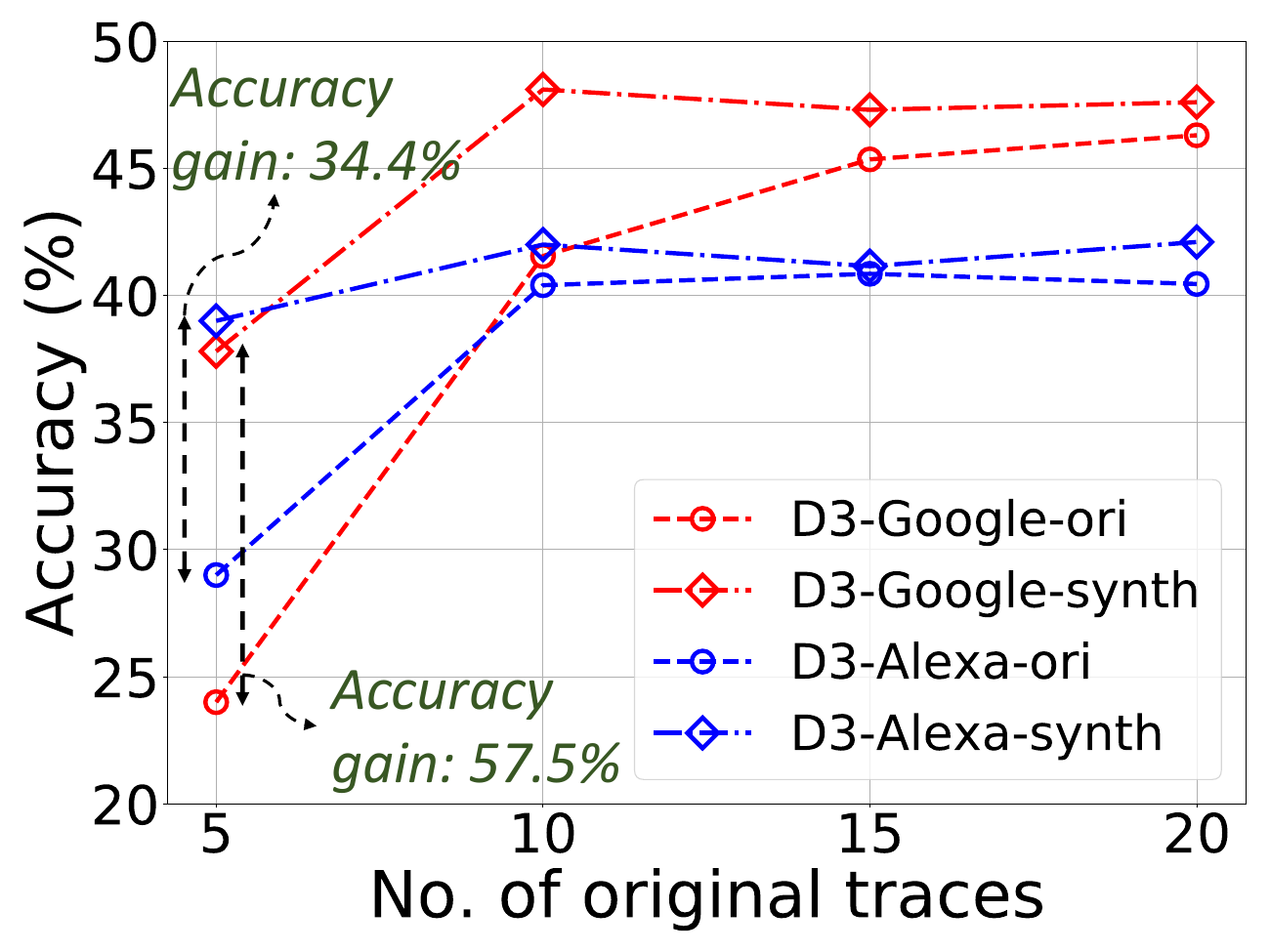}
  \vspace{-6mm}
  \caption{D3-Google and D3-Alexa}
  \label{fig:google-alexa}
\end{subfigure}
\vspace{-4mm}
\caption{Accuracy improvement with a limited number of original data for \textbf{original} and \textbf{synth} scenarios\vspace{-4mm}}
\label{fig:limited-original}
\end{figure}

A limited number of original traces is a challenging scenario which hinders the above ML performance. To see the support by \solution{} to improve the downstream ML accuracy, we change the number of original traces on \solution{} data generation and add the resulting synthetic traces to train the ML models. Fig.~\ref{fig:limited-original} reveals that when the number of original traces  is limited, \solution{} synthetic traces can exceed the accuracy of original data. In Fig.~\ref{fig:netflix-df}, \textbf{D2-DF }\textbf{synth} achieves 12.4\% accuracy gain and, in Fig.~\ref{fig:google-alexa}, \textbf{D3-Google} \textbf{synth} obtain 57.5\% accuracy gain compared to corresponding \textbf{original} scenarios. This is in contrast to the lower performances in \textbf{synth} scenario in Table~\ref{table:hierarchical clf} and highlights the advantages of \solution{} in data-limited use cases.

\subsubsection{Anomaly detection:} Anomaly detection often struggles with collecting sufficient  malicious data to train models. We extend the \textbf{D1} video fingerprinting at L3 while creating a class imbalance when training the ML models to mimic real world anomaly detection. We assume that randomly selected two classes with a limited number of traces are anomalies and another set of five classes with all available training traces are legitimate. 

We analyze two sub-cases. i)~\textit{sub-case 1}: Ground truth data is available for both legitimate and anomaly classes. In this case,  we simply calculate the accuracy of anomaly trace classification. To mimic the shorter duration and further generalize anomalous behaviour, we limit the trace length from 180s (Full trace) to first 45s.
ii)~\textit{sub-case 2}: Ground truth labels are available only for legitimate classes. During the test phase, we measure the uncertainty  of classification based on the entropy of classification results following a deep ensembling approach{~\cite{sensoy2018evidential}}. For the legitimate and anomaly traces, a lower and a higher uncertainty are expected respectively. 

\begin{figure}[t]
\centering
\begin{subfigure}{.49\columnwidth}
  \centering
  \includegraphics[width=\linewidth]{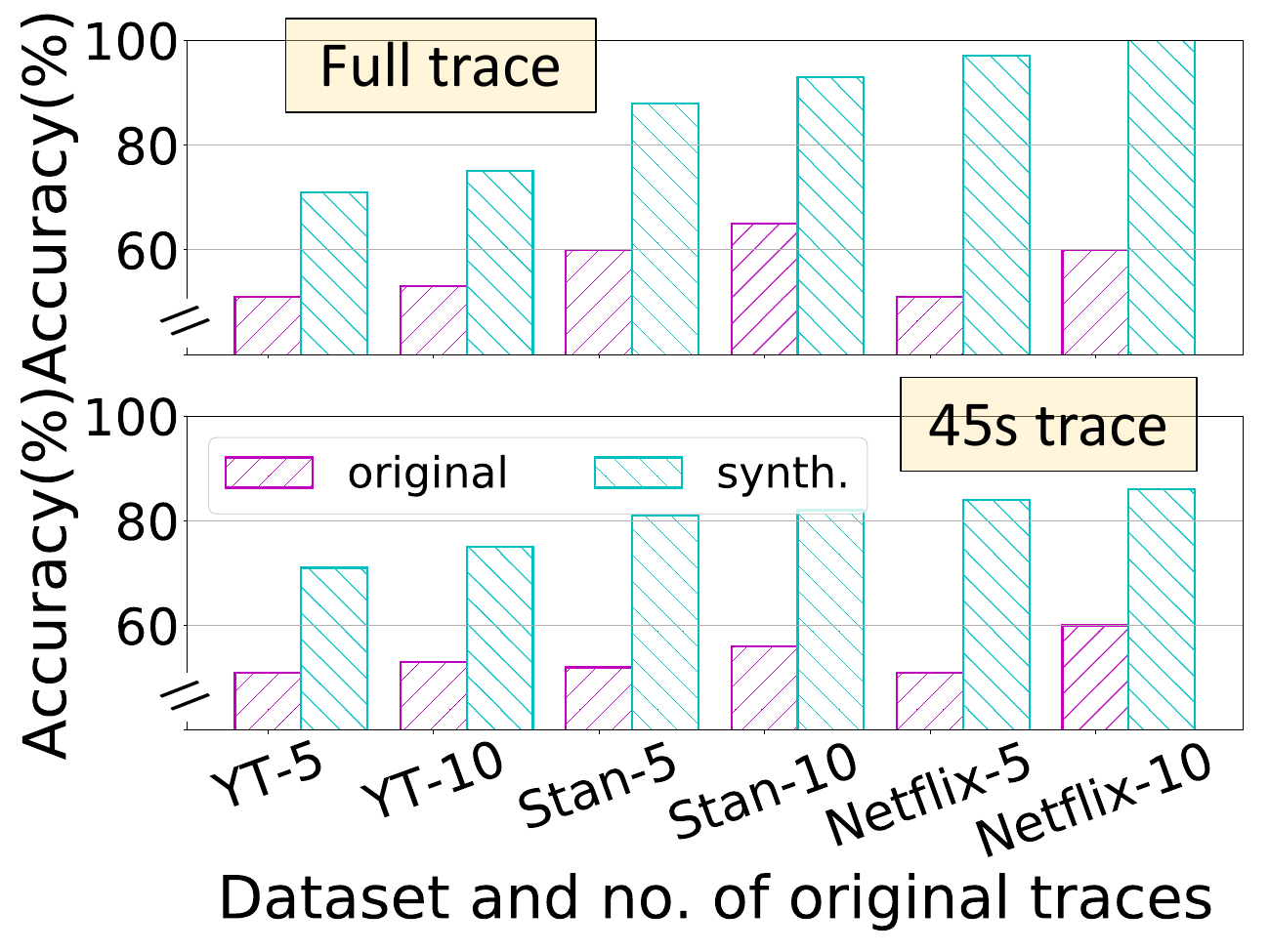}
  \vspace{-6mm}
  \caption{\textit{Sub-case~1}}
  \label{fig:anomaly-scenario-1}
\end{subfigure}%
\hfill
\begin{subfigure}{.49\columnwidth}
  \centering
  \includegraphics[width=\linewidth]{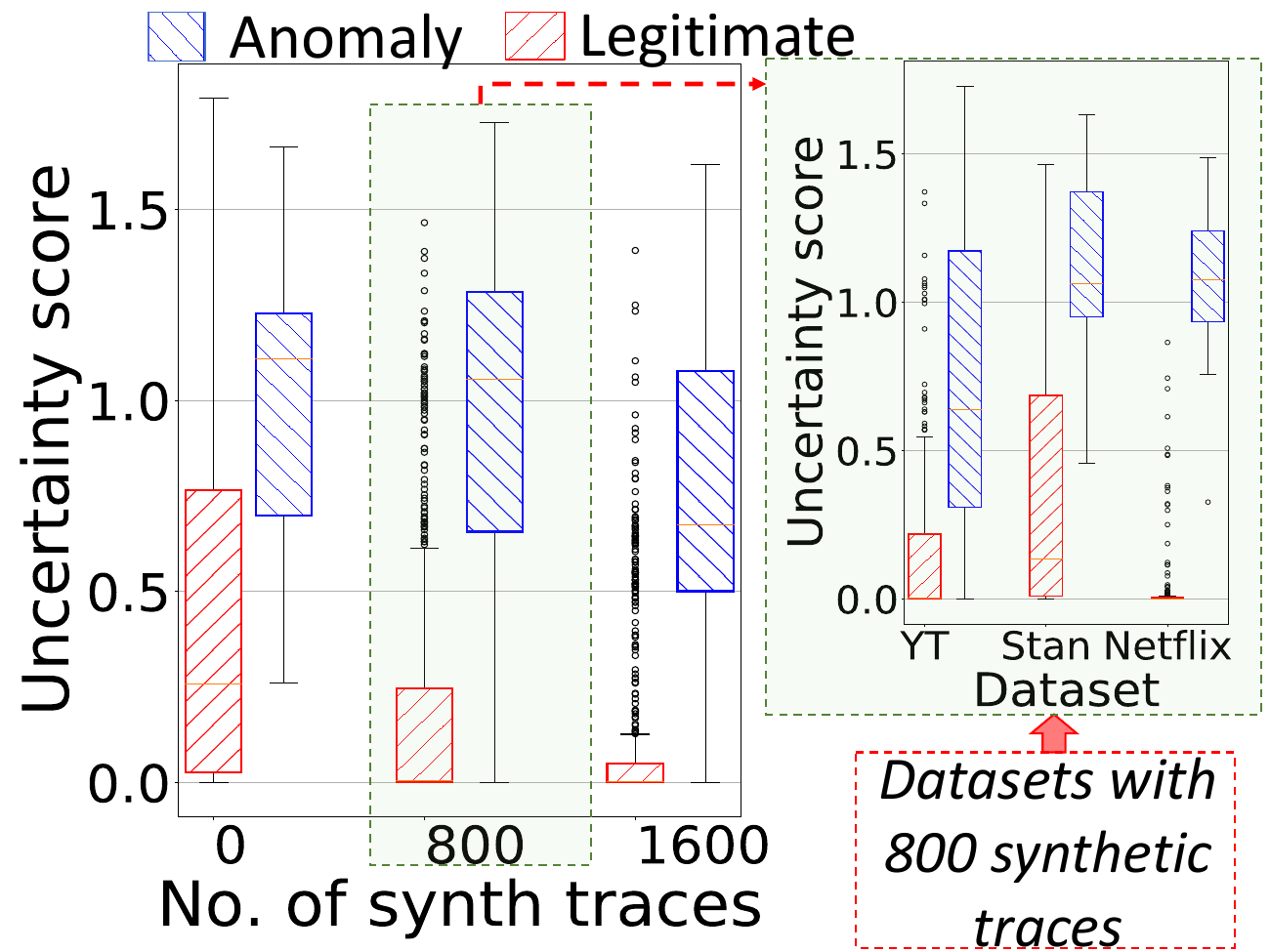}
  \vspace{-6mm}
  \caption{\textit{Sub-case~2}}
  \label{fig:anomaly-scenario-2}
\end{subfigure}
\vspace{-4mm}
\caption{Anomaly detection performance\vspace{-4mm}}
\label{fig:anomaly}
\end{figure}

\noindent
\textit{\textbf{Results and observation:}}
In \textit{sub-case~1} (Fig.~\ref{fig:anomaly-scenario-1}), for  \textit{Full} and 45s lengths, adding synthetic traces provide 54.6($\pm$18.3)\% and 48.5($\pm$9.0)\% of average gain respectively, compared to having only original data. We see that gain achieved is higher when the number of original data is limited. For example, \textbf{Netflix-5} achieves 90.2\% average accuracy gain compared to \textbf{Netflix-10} which has only 66.7\%. Fig.~\ref{fig:anomaly-scenario-2} presents the uncertainty comparison for legitimate and anomaly detection. In a gist, synthetic traces reduce the uncertainty of the predictions for legitimate samples (e.g., 0.75 on average at 1600 \textbf{synth} traces), whereas, the uncertainty for anomaly samples remains high. This higher uncertainty score is an indication to decide whether a given trace is an anomaly~\cite{sensoy2018evidential}.
This is further verified in the extended graph showing the score for different datasets at 800 \textbf{synth} data step. We notice a 0.74 and 1.01 average uncertainty difference between legitimate and anomaly data for \textbf{Stan} and \textbf{Netflix} respectively further evidencing the support of \solution{} for anomaly detection. 


\vspace{-1mm}
\subsubsection{Near real-time classification}

We analyze the \solution{} support for near real-time classification representing a scenario in which only a part of the network trace is extracted  without waiting for the entire trace. We assume that we can identify the beginning of the network trace. The corresponding GASF images are generated by cropping the initial GASF images from the bottom and right directions representing the 1D trace with limited data. We leverage L3 classification in Section~\ref{subsubsec:hie-clf}. We train multiple classifiers for different trace lengths which are measured in trace length for \textbf{D1} and percentage number of packets for \textbf{D2} and \textbf{D3}. 

\begin{figure}[t]
    \centering
    \includegraphics[width=0.9\columnwidth]{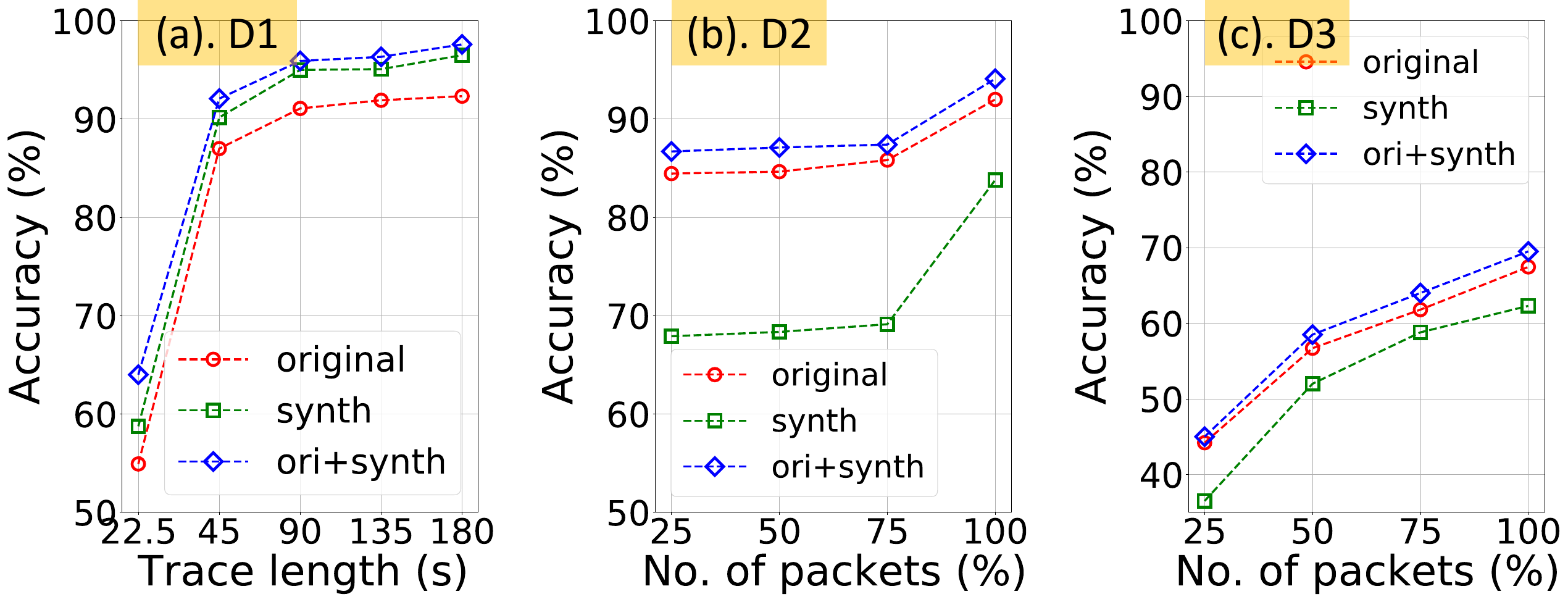} 
    \vspace{-4mm}
    \caption{Performance of L3 classification for different trace lengths/No. of packets.\vspace{-4mm}}
    \label{fig:crop_img}
\end{figure}

\textit{\textbf{Results and observation:}}
Fig.~\ref{fig:crop_img}(a) shows that with having only 45s worth of data, \textbf{D1} data can achieve over 92\% accuracy in \textbf{ori+synth} scenario which is 5.7\% accuracy gain compared to \textbf{original}. Though \textbf{synth} accuracy is less than the \textbf{original} scenario, on the one hand, \textbf{ori+synth} accuracy always outperforms \textbf{original} scenario in  \textbf{D2} and \textbf{D3} data in Fig.~\ref{fig:crop_img}(b) and Fig.~\ref{fig:crop_img}(c) respectively. On the other hand, \textbf{synth} accuracy follows the same increasing trend in \textbf{original} and \textbf{ori+synth}, and eventually, reducing the accuracy gap with \textbf{original}, for example in \textbf{D2}. Empirically, we observe GASF conversion takes time in the millisecond range (i.e., around 10 ms) without affecting overall inference process.

\vspace{-4mm}
\subsection{Comparison with 1D DM} \label{subsec:1D DM}
We compare L3 classification for \textbf{D1-Netflix} and \textbf{D3-Google} between synthetic data from 1D and 2D DMs from \solution{}. We follow, {a similar architecture to \solution{} DM when developing corresponding 1D DM and use the 1D traces to train them.} Once the 1D synthetic traces are generated,  we convert them to 2D GASF images for downstream ML classification. Table~\ref{table:1d vs 2d} reports that in all scenarios, synthetic data from 2D DMs in \solution{} overpass the data from 1D DMs and achieves almost the  same \textbf{original} accuracy. For instance,  in \textbf{synth} scenario, \solution{} data achieve 37.8\% and 30.2\% accuracy improvement over 1D DMs data for \textbf{D1-Netflix} and \textbf{D3-Google} respectively. We further measure a lower FID score for 2D DMs data which is 77\% and 36\% less than synthetic data from 1D DM in \textbf{D1-Netflix} and \textbf{D3-Google} respectively. These results reveal that 2D DMs in \solution{} can perform better than its 1D counterpart.

\vspace{-4mm}
\begin{table}[h!]
    \centering
    \captionsetup{justification=centering}
    \small{
    \caption{Comparison between 1D vs 2D DMs}
    \vspace{-4mm}
    \label{table:1d vs 2d}
    \begin{tabular}{lSSSSS}
        \hline
        \multirow{2}{*}{Dataset}&
            \multicolumn{1}{c}{\textbf{original} (\%)} &
            \multicolumn{2}{c}{\textbf{synth} (\%)} &
            \multicolumn{2}{c}{\textbf{ori+synth} (\%)} \\
            &  & {1D} & {2D} & {1D} & {2D} \\
        \hline
        \textbf{D1-Netflix} & 97.5 & 62.2 & 100.0 & 94.8 & 100.0  \\
        \textbf{D3-Google} & 74.0 & 42.3 & 72.5 & 73.2 & 76.0  \\
        \hline
    \end{tabular}
    }
\end{table}




\vspace{-6mm}
\section{Conclusion}
We presented \solution{}, a diffusion model (DM) based network traffic generation tool. It converts time-series network traffic data into a specific image format called Grammian Angular Summation Field (GASF). While addressing multiple challenges related to synthetic network traffic generation to achieve higher data fidelity, we also demonstrate the effectiveness of synthetic data in GASF format in various downstream ML tasks to improve their classification performance. Furthermore, we show that \solution{} exceeds the performance of SOTA GAN-based approaches and 1D DMs.

Although we have considered only single-variate time series, we can easily extend \solution{} for multi-variate time series and also accommodate metadata with a suitable mapping to integer values, by stacking up channels to a single image. At this stage, \solution{} deals with fixed lower image size to reduce the DM training time and to improve the scalability. To add more  variability in image sizes to represent different 1D trace lengths, we aim to leverage the stable/latent diffusion models which use lower dimensional embeddings of the input data~\cite{blattmann2023align, schramowski2023safe}. This can improve the scalability as well as the fidelity of the data. Privacy is yet another aspect we have not evaluated at this stage of \solution{}. We plan to introduce differentially private noise in the denoising process of DMs to synthesise privacy-aware datasets.


    


\appendix
\section*{Appendices}
\section{GASF conversion in detail}\label{append:gasf conversion}

First, a ${X}$, a normalized 1D signal (i.e., value range is between 0--1) where $X = {x_1, x_2, x_3,...,x_n}$ and $n$ is the number of samples, will be converted to polar coordinates as in Eq.~(\ref{eq:polar angle}) and Eq.~(\ref{eq:polar radius}).

\begin{equation}\label{eq:polar angle}
    \theta_i = \arccos{({x_i})}, ~ {x_i}\in {X}
\end{equation}
\vspace{-4mm}
\begin{equation}\label{eq:polar radius}
    rad_i = \frac{t_i}{C}
\end{equation}

\noindent
where, $t_i$ is the timestamp of $i^{th}$ sample and $C$ is a constant factor to regularize the radius. 

Then, in the polar coordinate system, the angular perspective can be easily exploited by considering the trigonometric sum between each point to identify the temporal correlation within different time intervals. Eq.~\ref{eq:gasf gen} shows how the GASF is formed. Here $I$ denotes the unit row vector, $[1, 1, ..., 1]$, $X^{'}$ is the transpose of $X$ and $i,j={1,2,3...n}$


\vspace{-2mm}
\begin{equation}\label{eq:gasf gen}
\begin{split}
    GASF    & = [\cos(\theta_i+\theta_j)]\\
            & = {X}^{'}.{X} - {\sqrt{I-{X}^2}}^{'}.{\sqrt{I-{X}^2}}
\end{split}
\end{equation}

\noindent
Further, setting [0,1] as the sample range of GASF images will add a bijective property to GASF images and hence, the time series can be traced back from the image.

\section{Improvement in ML by imaging network traces}\label{append:1d vs 2d original}

Table~\ref{table:1D vs 2D only original} reports the classification accuracy for \textbf{D1} dataset comparing original 1D traces and its corresponding GASF conversion. 1D and 2D classification models have the same architecture except the 1D and 2D convolutional layers in the respective models.
For all proportions of training data, 2D GASF exceeds the 1D time series accuracy by 1.5--13\% showing the improved performance  by GASF conversion. 

\begin{table}[h]
    \centering
    \small{
    \caption{L3 classification for D1 data using original 1D traces and corresponding GASF images}
    \vspace{-2mm}
    \label{table:1D vs 2D only original}
    \begin{tabular}{lcccccc}

        \hline
        {Platform} &
            \multicolumn{2}{c}{80\% data} &
            \multicolumn{2}{c}{40\% data} &
            \multicolumn{2}{c}{20\% data} \\
            \textbf{D1-} & 1D & 2D & 1D & 2D & 1D & 2D \\
             
        \hline
        
        \textbf{YT}  & 90.5 & \textbf{92.5} & 85.5 & \textbf{87.0} & 71.0 & \textbf{75.0}  \\
        \textbf{Stan}  & 91.0 & \textbf{99.0} & 96.0 & \textbf{98.0} & 90.0 & \textbf{93.0}\\
        \textbf{Netflix}  & 83.0 & \textbf{100.0} & 84.0 & \textbf{97.0} & 78.0 & \textbf{88.0}  \\
        \hline
    \end{tabular}
    }
\end{table}

\section{Further analysis}

\subsection{Histograms of D3 original and synth}\label{append:histo d3}

\begin{figure}[h]
    \centering
    \begin{subfigure}{.48\columnwidth}
        \centering
        \includegraphics[width=\linewidth]{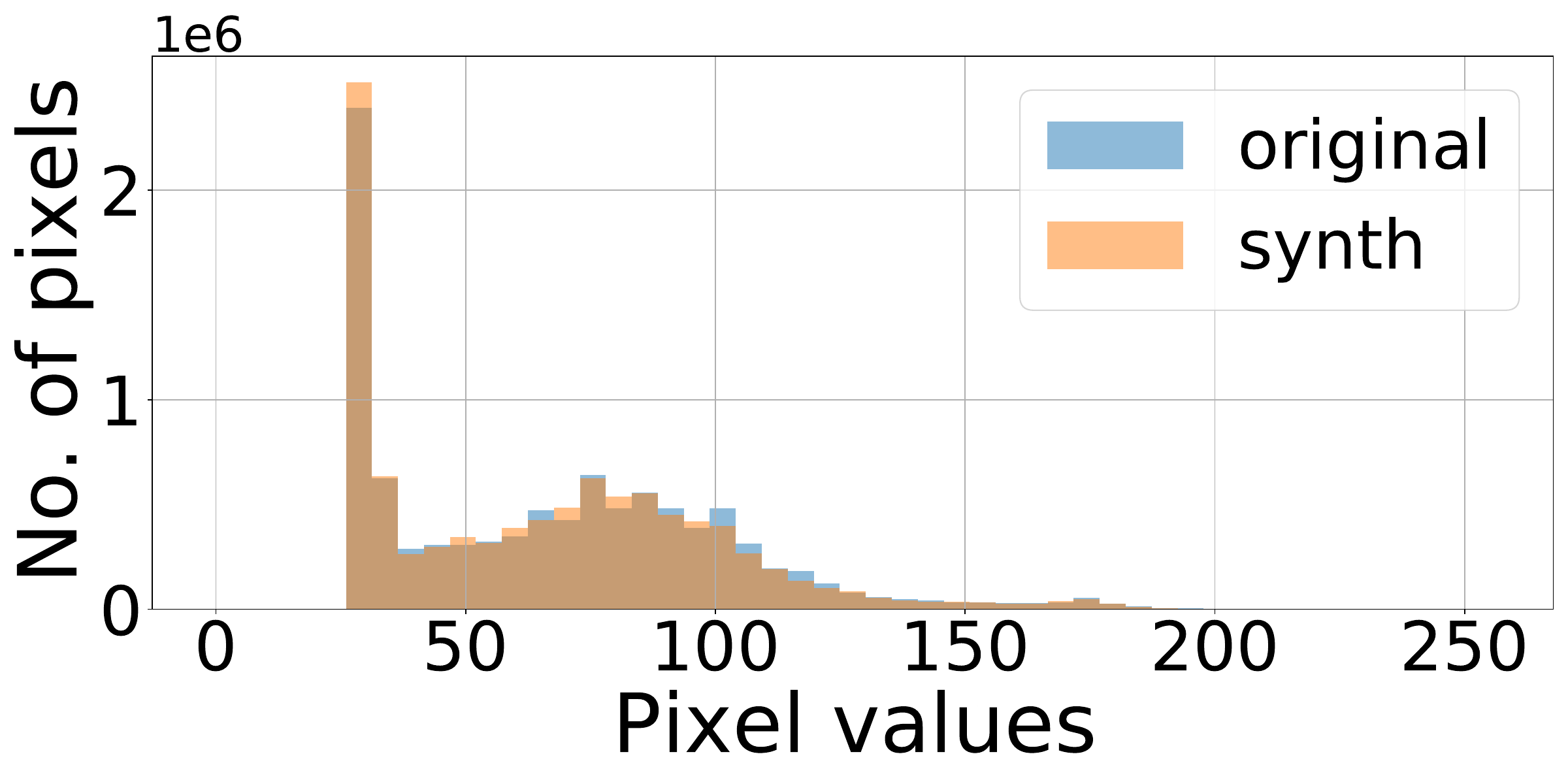}
        \vspace{-6mm}
        \caption{D1-(YT, Stan, Netflix)}
        \label{fig:histo-d1}
    \end{subfigure}
    \hfill
    \begin{subfigure}{.48\columnwidth}
        \centering
        \includegraphics[width=\linewidth]{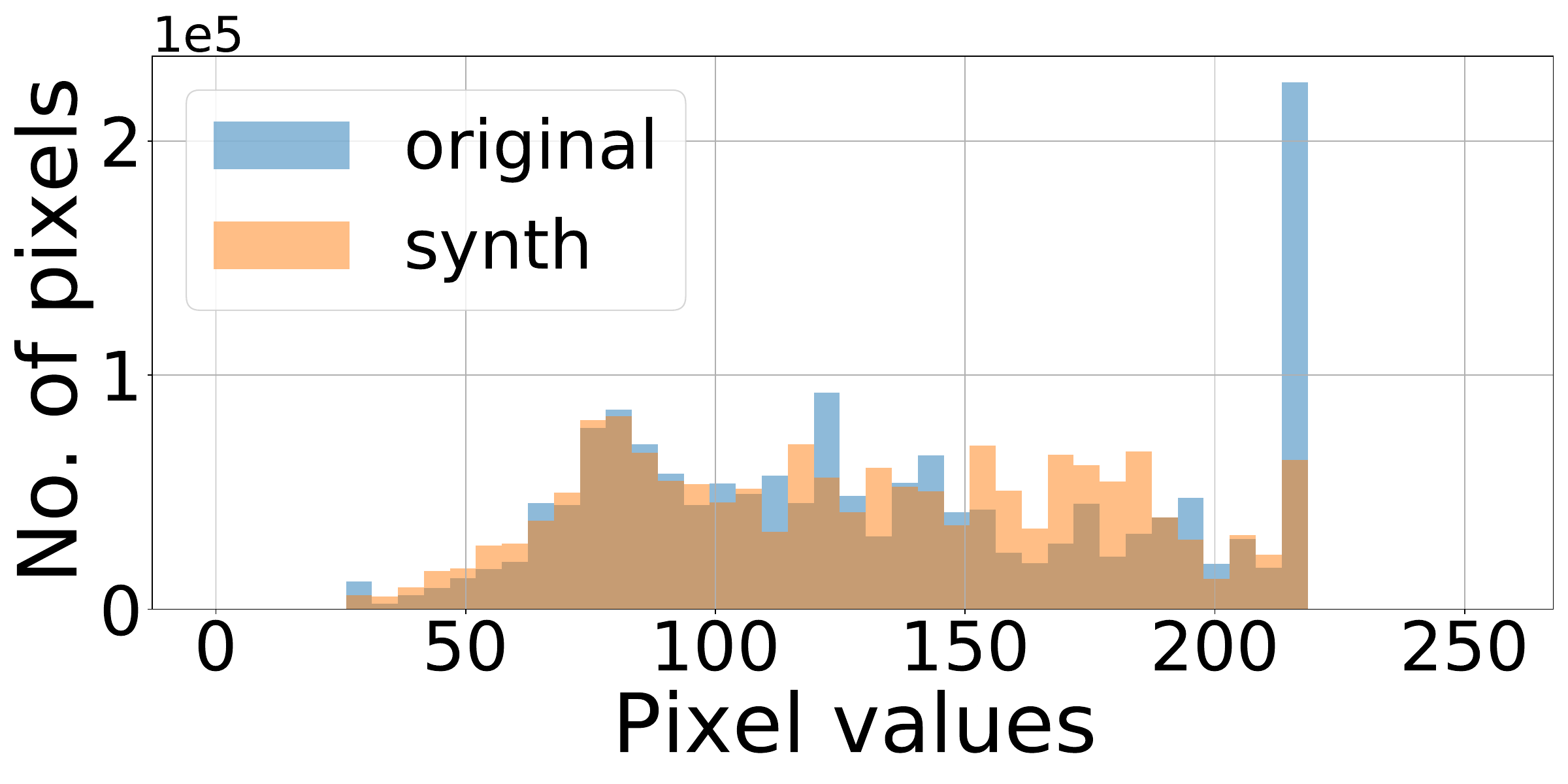}
        \vspace{-6mm}
        \caption{D2-DF}
        \label{fig:hist-d2}
    \end{subfigure}
    \hfill
   \begin{subfigure}{.48\columnwidth}
        \centering
        \includegraphics[width=\linewidth]{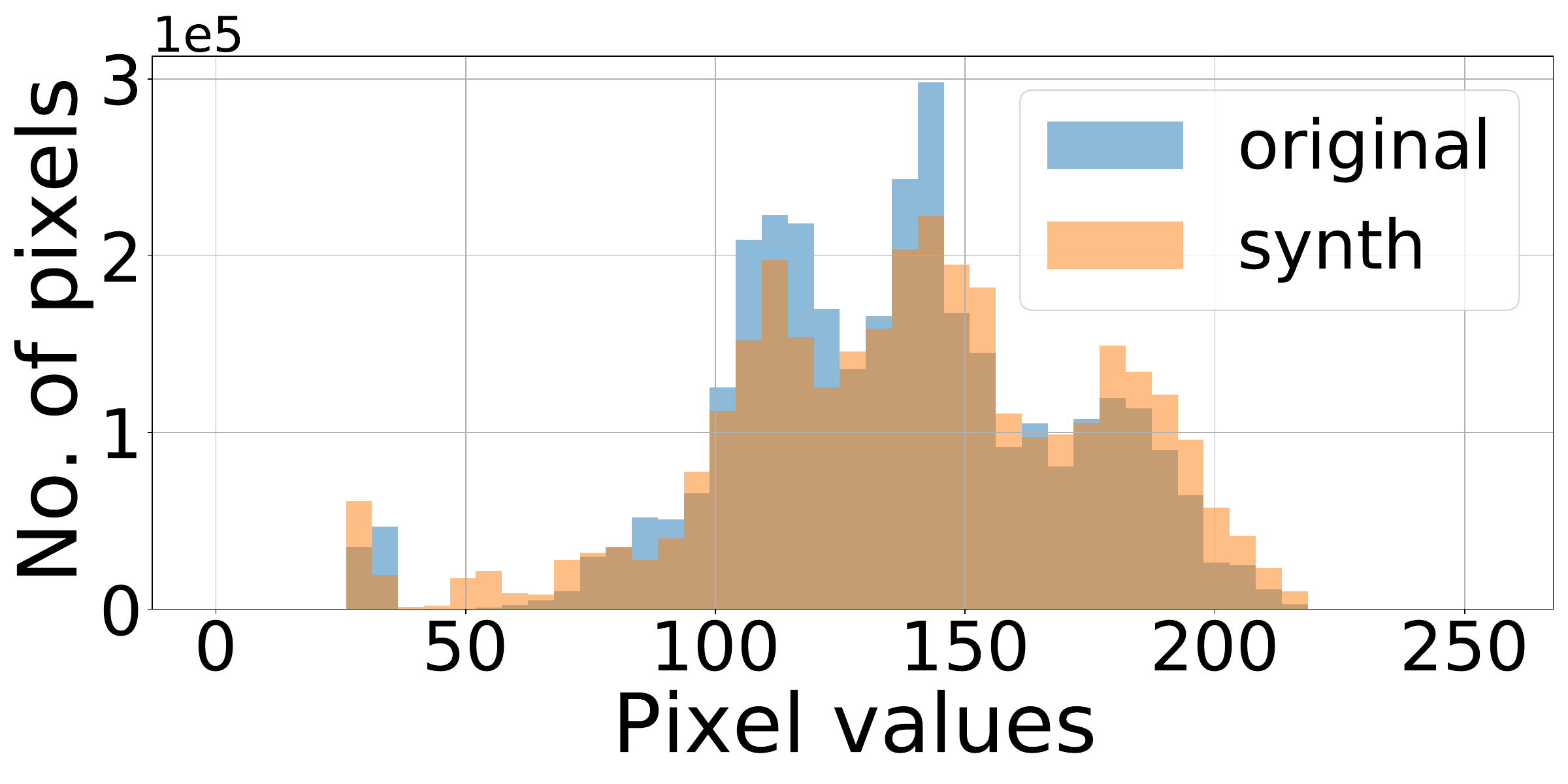}
        \caption{D3-Google}
        \label{fig:histo-d3-google}
    \end{subfigure}
    \hfill
    \begin{subfigure}{.48\columnwidth}
        \centering
        \includegraphics[width=\linewidth]{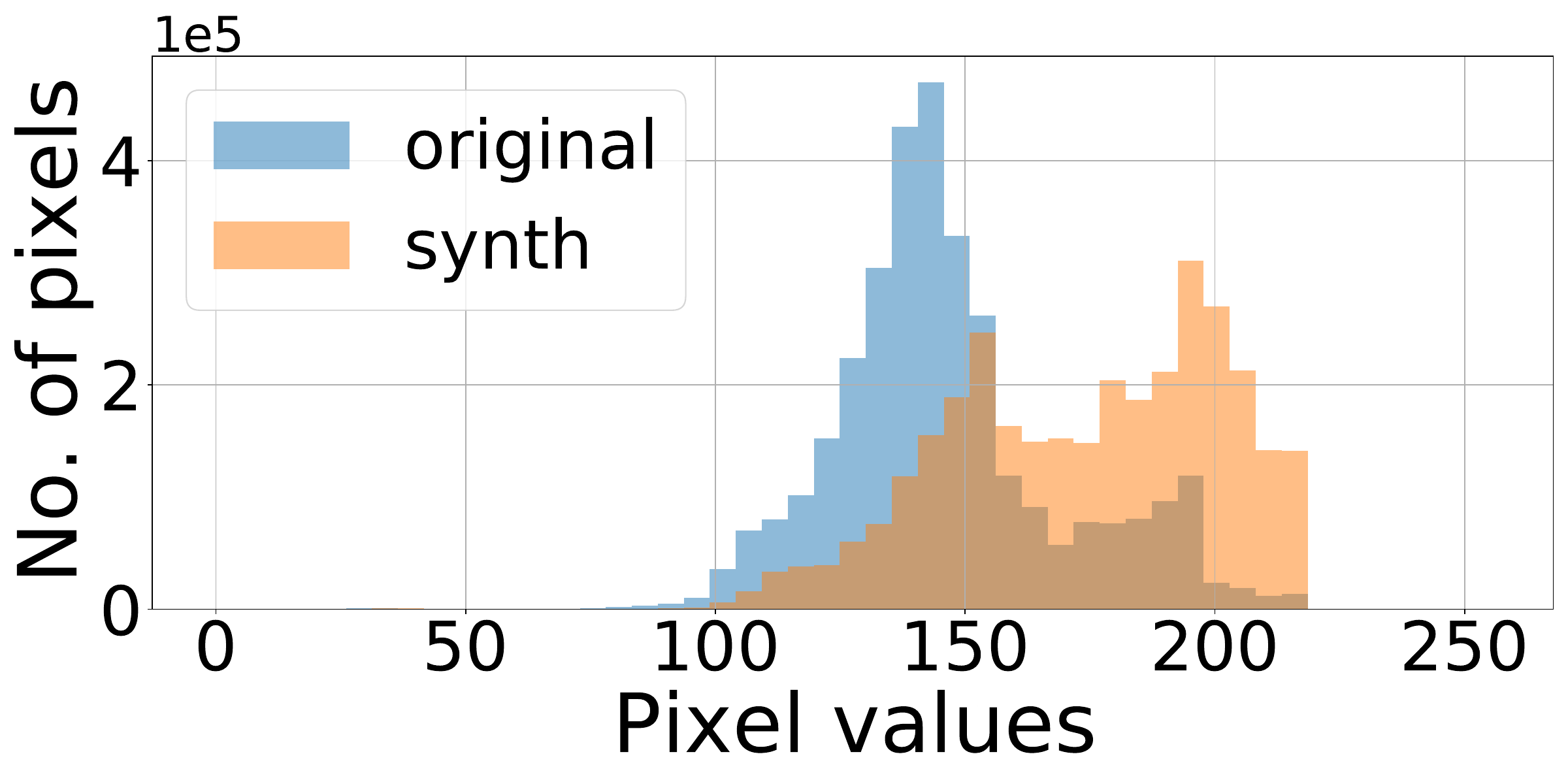}
        \caption{D3-Alexa}
        \label{fig:histo-d3-alexa}
    \end{subfigure}
\vspace{-4mm}
\caption{Histogram distribution between original and \solution{} synthetic data}
\label{fig:histogram}
\end{figure}

We compare the histograms between \textbf{original} and \textbf{synth} GASF images from \textbf{D1}, \textbf{D2} and \textbf{D3} in Fig.~\ref{fig:histogram}. Overall, original and synthetic data have a similar distribution preserving a wide range of pixel values.  We observe a noticeable difference in the distribution of \textbf{D3-Alexa} in Fig.~\ref{fig:histo-d3-alexa}. This is  due to the higher similarity between the classes that has hindered DMs learning unique differences between \textbf{D3-Alexa} classes. These high overlaps between histograms verify that attributes such as packet size, and temporal correlations that were mapped onto the original GASF images are still maintained in \solution{} synthetic data.

\subsection{Impact of adding synthetic data}\label{append:impact of adding synth}

\begin{figure}[h]
\centering
\begin{subfigure}{.48\columnwidth}
  \centering
  \includegraphics[width=\linewidth]{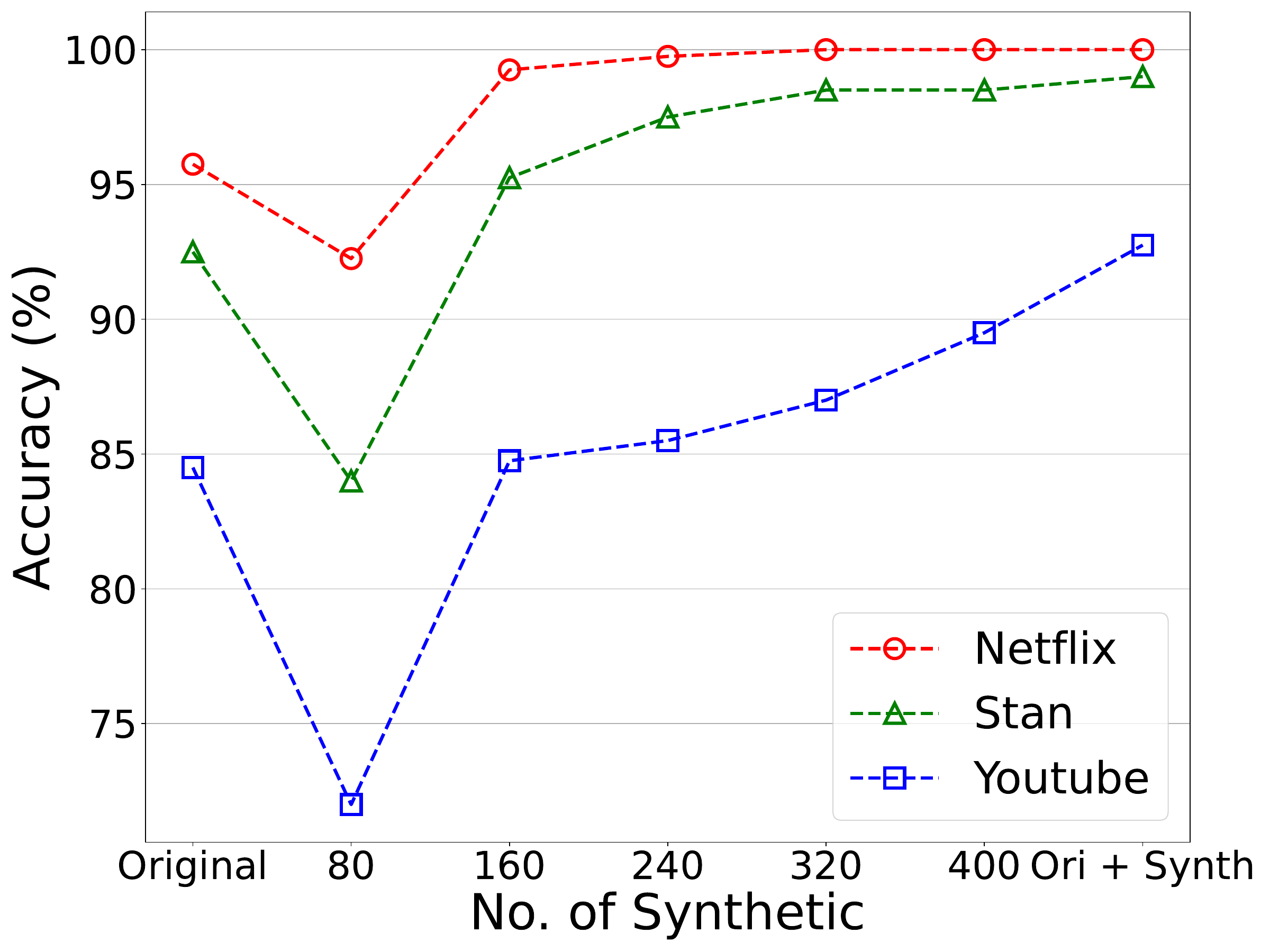}
  \caption{D1-(YT, Stan, Netflix)}
  \label{fig:synth d1}
\end{subfigure}%
\hfill
\begin{subfigure}{.48\columnwidth}
  \centering
  \includegraphics[width=\linewidth]{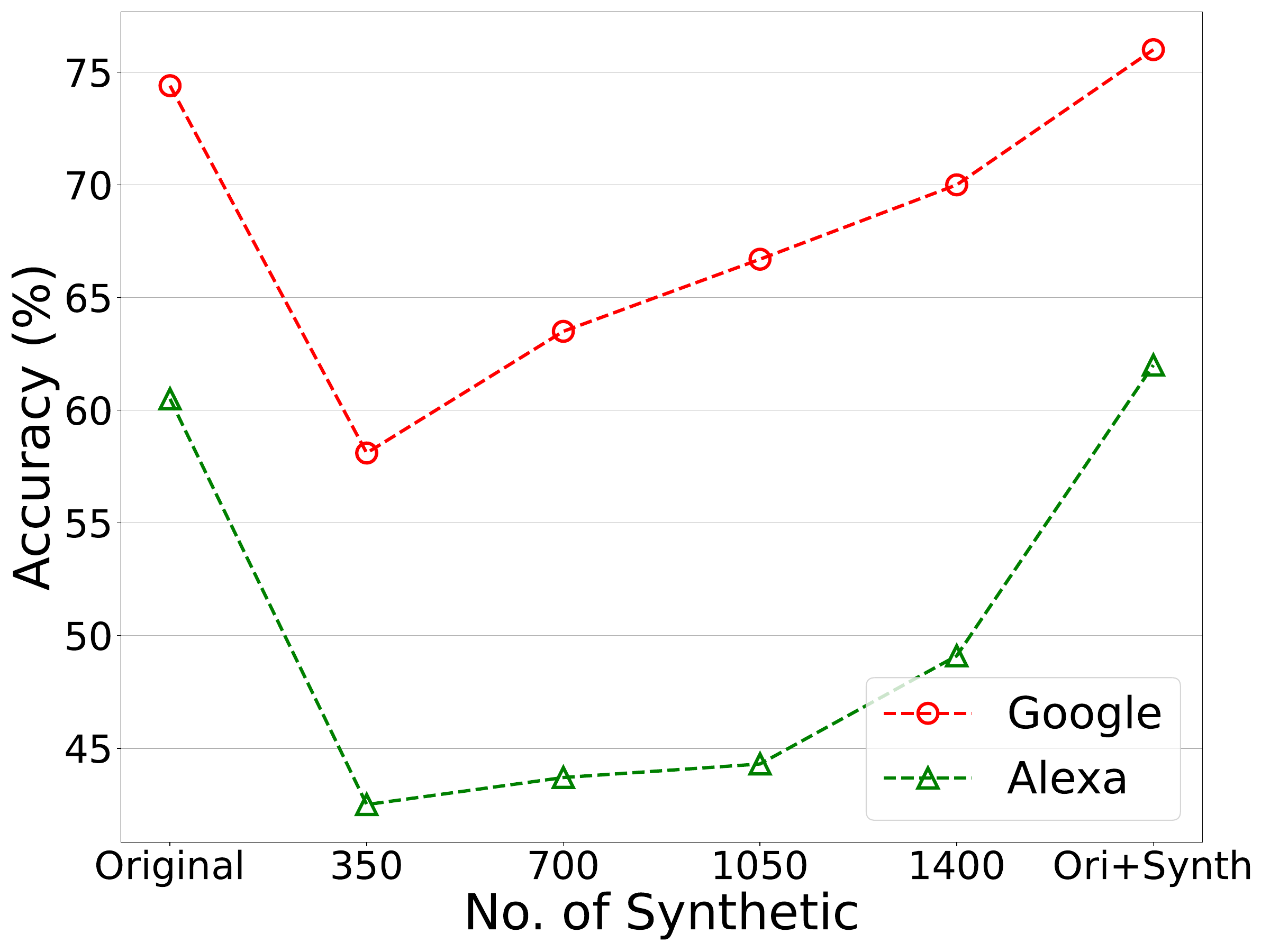}
  \caption{D3-(Google, Alexa)}
  \label{fig:synth d3}
\end{subfigure}
\caption{Impact of number of synthetic data on L3 classification accuracy}
\label{fig:adding synth}
\end{figure}

Fig.~\ref{fig:adding synth} illustrates the accuracy variation when gradually adding the synthetic data. We start from the \textbf{original} scenario and the numerical values in the x-axis show the number of synthetic images added in \textbf{synth} scenario (i.e., having only synthetic images). Last index in the x-axis indicates the \textbf{ori+synth} scenario. Fig.~\ref{fig:synth d1} shows that even with 160 synthetic data all three datasets in \textbf{D1} can surpass or achieve the same accuracy of \textbf{original } data. Though we observe a sudden drop in \textbf{synth} scenario for \textbf{D3} data, adding synthetic data shows a gradual increase in accuracy for \textbf{D3} data in Fig.~\ref{fig:synth d3}, and eventually exceeds the \textbf{original} accuracy in \textbf{ori+synth} scenario.

\balance
\bibliographystyle{ACM-Reference-Format}
\bibliography{sample-base}

\end{document}